\documentstyle[11pt,aasms4,rotating]{article}

\def\lsun{{\,L_\odot}}
\def\simlt{\lower.5ex\hbox{$\; \buildrel < \over \sim \;$}}
\def\simgt{\lower.5ex\hbox{$\; \buildrel > \over \sim \;$}}

\def\cm{{\rm\,cm}}
\def\km{{\rm\,km}}

\def\pc{{\rm\,pc}}

\def\sec{{\rm\,s}}

\def\gcm3{{\rm\,g\,cm^{-3}}}
\def\ncm3{{\rm\,cm^{-3}}}
\def\kelvin{{\rm\,K}}
\def\erg{{\rm\,erg}}

\def\>{$>$}
\def\<{$<$}

\def\refindent{\par\noindent\hangindent=3pc\hangafter=1 }
\def\aa#1#2#3{\refindent#1, A\&A, {\bf#2}, #3.}
\def\aalett#1#2#3{\refindent#1, A\&A {\it (Letters)}, {\bf#2}, #3.}

\def\apj#1#2#3{\refindent#1, {\it ApJ}, {\bf#2}, #3.}
\def\apjlett#1#2#3{\refindent#1, {\it ApJ (Letters)}, {\bf #2}, #3.}

\def\nature#1#2#3{\refindent#1, {\it Nature}, {\bf #2}, #3.}

\lefthead{Fatuzzo et al.}
\righthead{Electron-positron Annihilation Radiation from Sgr A East at the Galactic Center}

\begin{document}
\centerline{Submitted to the Editor of the Astrophysical Journal.}
\vskip 0.5in
\title{Electron-positron Annihilation Radiation\\
       from Sgr A East at the Galactic Center}

\author{Marco Fatuzzo$^{1,2}$, Fulvio Melia$^{2,3,}$\altaffilmark{4}, and
Johann Rafelski$^2$}
\affil{$^1$Physics Department, Xavier University, Cincinnati, OH  45207}
\affil{$^2$Physics Department, The University of Arizona, Tucson, AZ 85721}
\affil{$^3$Steward Observatory, The University of Arizona, Tucson, AZ 85721}


\altaffiltext{4}{Sir Thomas Lyle Fellow and Miegunyah Fellow.}


\begin{abstract}
Maps of the Galactic electron-positron annihilation radiation show
evidence for three distinct and significant features: (1) a central
bulge source, (2) emission in the Galactic plane, and (3) an
enhancement of emission at positive latitudes above the Galactic
Center.  In this paper, we explore the possibility that Sgr A East,
a very prominent radio structure surrounding the Galactic nucleus,
may be a significant contributer to the
central bulge feature.  The motivation for doing so stems
from a recently proposed link between this radio object and the EGRET
$\gamma$-ray source 2EG J1746-2852. If this association is correct, then Sgr A East
is also expected to be a source of copious positron production.  The results presented
here show that indeed Sgr A East must have produced a numerically significant
population of positrons, but also that most of them have not yet had
sufficient time to thermalize and annihilate.  As such, Sgr A East
by itself does not appear to be the dominant {\it current} source of
annihilation radiation, but it will be when the positrons have
cooled sufficiently and they have become thermalized.  This raises the 
interesting possibility that the bulge component may be due to the
relics of earlier explosive events like the one that produced Sgr A East.
\end{abstract}




%

\section{Introduction}
Electron-positron annihilation radiation from the Galactic Center region
was first reported in the 1970s after the successful flight of a series
of balloon-borne instruments (Johnson, Harnden \& Haymes 1972;
Johnson \& Haymes 1973;  Haymes et al. 1975).  An unambiguous identification
of the 511 keV line and the detection of the positronium continuum
followed several years later (Leventhal, MacCallum \& Stang 1978).  Since
that time, numerous follow-up experiments have provided additional
monitoring of this source, though not always with a consistent
measurement of the location or distribution of the emission, nor on the
possible time variability of the 511 keV line flux (see the review by
Tueller 1993). However, the Oriented Scintillation Spectrometer
Experiment (OSSE) on the {\it Compton Gamma-Ray Observatory} was recently
able to carry out high-sensitivity observations of the Galactic plane
and uniquely map the distribution of positron annihilation radiation and search
for time variability of the emission (Purcell et al. 1997).

The resulting maps show evidence for three distinct and significant
features: (1) a central bulge source with a total flux (line plus 
continuum) of $\sim 3.3\times 10^{-4}$ photons cm$^{-2}$ s$^{-1}$ 
and a Full Width at Half Maximum (FWHM) of
$\sim 4$ degrees, (2) emission in the Galactic plane with a total
flux of $\sim 10^{-3}$ photons cm$^{-2}$ s$^{-1}$ and a broad extension
of more than $9$ degrees FWHM in both latitude and longitude, and
(3) an enhancement or extension of emission at positive latitudes above
the Galactic Center with a total flux of $\sim 9\times 10^{-4}$ photons 
cm$^{-2}$ s$^{-1}$ and a FWHM of over $\sim 16$ degrees.  
The average spectrum for the 1991 December viewing period of the 
Galactic Center (central bulge plus enhancement) exhibits a clear
line feature at $509\pm5$ keV.  The line width ($2.5$ keV) is consistent with
the instrumental resolution.  The observed ratio of the 511 keV
line flux to the total positronium flux corresponds to a positronium
fraction of $0.98\pm0.04$.  In addition, OSSE found
no evidence for time variability.

Sources proposed as contributers to the annihilation 
emission have included cosmic-ray interactions in the interstellar
medium (Lingenfelter \& Ramaty 1982), pulsars (Sturrock 1971)
and $\beta^+$ decay products from radioactive nuclei (e.g.,
$^{56}$Co, $^{44}$Sc, and $^{26}$Al) produced by supernovae, novae,
or Wolf-Rayet stars (Clayton 1973; Ramaty \& Lingenfelter
1979; Signore \& Vedrenne 1988; Woosley \& Pinto 1988;
Lingenfelter \& Ramaty 1989).  However, efforts to
model the distribution of the 511 keV emission have been only
moderately successful (see, e.g., Purcell et al. 1993).  These models,
consisting of components generally representing Galactic emission
observed at other wavelengths (e.g., CO maps) or geometric
configurations assuming that the emission of annihilation radiation
follows a Galactic distribution of known sources, were not able
to describe the OSSE maps adequately.  It is evident that simple
Galactic distributions are not sufficient to describe the most
recent 511 keV data and that additional components are required. 
For example, the expected 511 keV line flux resulting from
the radioactive decay of $^{26}$Al represents only $\sim
3-9\%$ of the total Galactic 511 keV line flux (Purcell et al. 1997).

In this paper, we explore the possibility that Sgr A East,
a very prominent elongated radio structure surrounding 
(although slightly off-centered from) the Galactic nucleus, is a significant
source of the annihilation radiation associated with the bulge 
feature.  The motivation for doing so stems from the 
association of this radio object and the EGRET $\gamma$-ray source
2EG J1746-2852 suggested by Melia et al. (1998).  These authors show that
the high-energy component of the EGRET spectrum can be accounted
for by the decay of neutral pions ($\pi^0\rightarrow\gamma\gamma$)
produced in $p-p$ scatterings between ambient protons and a
relativistic population accelerated by shocks within the Sgr A
East shell.  Since charged pions which decay leptonically ($\pi^\pm\rightarrow
\mu^\pm\nu_\mu$, with $\mu^\pm\rightarrow e^\pm\nu_e\nu_\mu$) are
also produced,  the rest of Sgr A East's spectrum, extending down to
GHz energies, is filled uniquely by the bremsstrahlung, Compton and
synchrotron interactions of the associated electrons and positrons.
The VLA data, for example, 
are consistent with the synchrotron emissivity of the decay leptons within
an equipartition magnetic field of $\approx 10^{-5}$ G.

If this scenario is correct, Sgr A East must be a site of copious
positron production.  Here, we investigate whether this process is
sufficient to power the measured flux of annihilation radiation
from the Galactic Center, and if not, whether theses observations
suggest an evolutionary history of Sgr A East.

The rest of the paper is organized as follows.
Section 2 presents what is known about
Sgr A East from observations as well as what we infer based on the results of
Melia et al. (1998).  This section also includes a calculation of the ionization
state of the gas in Sgr A East.  The fate of positrons produced in Sgr
A East is then considered in Section 3.  Two different
evolutionary cases for this environment are presented in Section 4.
The characteristics of the annihilation radiation spectra for these cases
are determined in Section 5 and compared to observations.  
Our concluding remarks are also given in Section 5.

\section{The Environment of Sgr A East}

\subsection{The Basic Properties}

Sgr A East is an elliptical structure with nonthermal radio emission
peaked at its periphery.  Geometrically, the structure is elongated
along the Galactic plane with a major axis of length 10.5 pc and a
center displaced from the apparent dynamical nucleus, Sgr A West, by
2.5 pc in projection toward negative Galactic latitudes.  The actual
distance between Sgr A West and the geometric center of Sgr A East
has been estimated to be $\sim 7$ pc  (Yusef-Zadeh \& 
Morris 1987; Pedlar et al. 1989; Yusef-Zadeh et al. 1999).
Although morphologically similar to supernova remnants (SNRs),
the energetics ($\sim 4\times 10^{52}$ ergs) of Sgr A East,
based on the power required to carve out the radio synchrotron remnant within
the surrounding dense molecular cloud, appear to be extreme compared to
the total energy ($\sim 10^{51}$ ergs) released in a typical
supernova (SN) explosion.  It has been suggested that Sgr A East may be
the remnant of a tidally disrupted star (Khokhlov \& Melia 1996).
Regardless of what actually caused the initial
explosion, recent observations of this region at 1720 MHz (the transition
frequency of OH maser emission) have revealed the presence of several 
maser spots at the SE boundary of Sgr A East with a velocity of $\approx 50
\;\km\;\sec^{-1}$, and one near the Northern arm of Sgr A West at a velocity
of $134 \;\km\;\sec^{-1}$ (Yusef-Zadeh et al. 1996).  These observations
are consistent with the presence of shocks produced at the interface between
the expanding Sgr A East shell and the surrounding environment.  The
implied age of Sgr A East ($\tau\sim 5\pc / 100 \;\km\;\sec^{-1}\sim 
5\times 10^4$ yr) is then consistent with those of typical SNRs.

Given its proximity to the Galactic Center, 
Sgr A East is bathed by intense IR and UV radiation fields
associated with the central 1 $\sim$ 2 parsecs of the galaxy
(Telesco et al. 1998; Davidson et al. 1992; Becklin, Gatley \&
Werner 1982; Melia, Yusef-Zadeh \& Fatuzzo 1998).  
Although there are no direct measurements of the properties
of the ambient medium enveloped by Sgr A East, X-ray observations suggest
an average electron ambient density of $\sim 3 - 6\;\cm^{-3}$ (Koyama et al. 1996; 
Sidoli \& Mereghetti 1999).  These results are consistent with an upper limit of $\sim
30\;\cm^{-3}$ inferred by Melia et al. (1998). The temperature of the ambient
medium is considerably less constrained.
Observationally, the ISM within the inner 150 pc is comprised of
several gas components with a range of temperatures.  Inside the minicavity,
the electron temperature lies in the range $4000-7000$ K, and is comparable
to the average value of $7000$ K found for Sgr A West (Roberts,
Yusef-Zadeh \& Goss 1996). However, on a larger scale (i.e., tens of
pc), the X-ray spectra observed with ASCA exhibit several emission 
lines from highly
ionized elements that are characteristic of a $\sim 10$ keV thermal
plasma.  In particular, this region stands out in the intense
$6.7$ keV K$\alpha$-transition of He-like Fe (Yamauchi, et al. 1990;
Koyama, et al. 1996).  These spectra indicate a common origin for
the plasma in the entire region, rather than a superposition of
individual sources.  Assuming an expansion velocity comparable
to the sound speed for $kT\sim 10$ keV, Koyama et al. (1996) infer
an expansion age of $\sim 10^5$ yr, which is similar to
that of Sgr A East.  Thus, both the X-ray plasma and the
relativistic synchrotron source may have been produced by
the same explosive event $\sim 0.5-1\times 10^5$ yr ago.  The
BeppoSAX observations of the Sgr A Complex suggest somewhat different
characteristics, in which the high-energy data are well fit by the sum of
two thermal models with $kT\sim 0.6$ and $8$ keV, with the lower 
temperature plasma well correlated to Sgr A East (Sidoli \& Mereghetti 1999).

\subsection{Other Inferred Characteristics}

Adopting the scenario of Melia et al. (1998) invoked to account for the
Galactic Center EGRET observations, we assume that pion-production
occurs within the Sgr A East shell as a result of collisions between
shock-accelerated and ambient protons (see also
Markoff, Melia \& Sarcevic 1997).  A crucial aspect of this
mechanism is the decay of charged pions to muons, and subsequently,
to relativistic electrons and positrons.  The rate at which these
leptons are produced is linked directly to the rate at which neutral 
pions are produced
(since the charged pions and neutral pions are both byproducts of the
same scattering events), and is therefore well constrained by the EGRET 
observations (Markoff, Melia \& Sarcevic 1999).  
In addition, since the relativistic leptons radiate 
synchrotron emission, their energy distribution is well constrained by
the radio properties of Sgr A East.  These observations also provide 
an indirect measure of the magnetic field strength in Sgr A East (see below).

In order to keep the analysis as concise as possible while at the same
time capturing the important characteristics of the Sgr A East environment, 
we assume a uniform and homogeneous shell geometry, with an inner radius of
4 pc and an outer radius of 5pc, and a centroid about $R = $ 7 pc from the
Galactic Center.  The IR and
UV radiation fields bathing this region are assumed to have blackbody 
spectra with temperatures $T_{ir} = 100$ K and
$T_{uv} = 30,000$ K, respectively, and energy densities
\begin{equation}
u = {L\over 4\pi R^2 c} = 4.4 \times 10^{-10}\;\erg\;\cm^{-3}
\;\left({L\over 2\times 10^7 \;\lsun}\right)
\;\left({R\over 7\pc}\right)^{-2}\;,
\end{equation}
where $L$ is either $L_{ir} = 10^7 \lsun$, or $L_{uv} = 2\times 10^7 \lsun$.
Furthermore, we will assume that Sgr A East has an age
of $75,000$ years and is comprised mostly of
hydrogen with an average density $n_H = 5\cm^{-3}$, and helium with 
an average density $n_{He} = 0.5\cm^{-3}$.  
The temperature profile $T$ within the Sgr A Complex must reflect the
superposition of the various components discussed above.  For our purposes
here, however, we assume that the ambient plasma where the positrons are produced 
may be ascribed a single characteristic value of $T$ at any given time.
The positrons are assumed to be produced within the Sgr A East
shell in accordance with Case 2 of Melia et al. (1998), the one that produced
the best fit to the gamma-ray data.  The injection function $I_+$
for these particles is shown in Figure 1 as a function of the positron
energy $E_+$.  The total production rate,
found by integrating this function over 
all values of $E_+$, is equal to $2.7\times 10^{-18}$ cm$^{-3}$ s$^{-1}$. 
The corresponding magnetic field strength, inferred from the VLA
spectrum, is $B = 1.1\times 10^{-5}$ G, which is within a factor of 2 of 
its equipartition value.  

One of the processes by which
annihilation occurs is charge exchange between positrons and atoms.
Our analysis of the electron-positron annihilation considered below
will therefore require an assessment of the ionization fractions 
in the Sgr A East shell, which are set by the balance of total
recombination and ionization rates for hydrogen and helium.  
For simplicity, we assume that all atoms are ionized from their ground state.
We will confirm {\it a posteriori} that the ambient plasma in Sgr A East
is highly ionized and that therefore the approximations we make here
do not greatly affect the results. 
We can therefore ignore the trace presence of neutral helium as the charge exchange 
process is dominated by the more prevelant singly ionized species of 
this atom.   
In addition, the number density of ambient electrons is well-approximated 
by the constant value $n_e = 6 \cm^{-3}$. The recombination rates 
are therefore independent of the ionization 
fractions of the species comprising the plasma, and hence the
ionization equations for hydrogen and helium decouple and reduce to
the following:
\begin{equation}
{n^{HI}\over n^{HII}} = {\sum_i R_{fi}^{HII}\over R_{1f}^{HI}
+ C_{1f}^{HI}}\;,
\end{equation}
\begin{equation}
{n^{HeII}\over n^{HeIII}} = {\sum_i R_{fi}^{HeIII}\over
R_{1f}^{HeII} + C_{1f}^{HeII}}\;,
\end{equation}
where $R_{fi}$ is the radiative recombination rate 
to state $i$ and $R_{1f}$ and $C_{1f}$ are the radiative and collisional ionization
rates from the ground state. Collisional recombination can be ignored for the 
conditions found in Sgr A East.
 
Since the electron self-collision rate 
\begin{equation}
R_{c} \sim {1/t_c} \sim 7.1\times 10^{-10} \sec^{-1}\; \ln \Lambda
\left({T\over 10^7 \kelvin}\right)^{-3/2}\left({n_e\over 6\cm^{-3}}\right)
\end{equation}
(Spitzer 1956) is much larger than the radiative recombination rates in the 
Sgr A East environment (see Figure 2), the free 
electrons thermalize prior to recombining.
We can therefore use the collisional ionization rates for HI and HeII in terms of the
ambient electron temperature $T$ and the 
corresponding ground-state ionization energy $\chi$
from Bell et al. (1983).  These
represent the best fits to the observation and theory, and take the form 
\begin{equation} C_{1f} = n_e \exp(-\chi/ kT)\;\left(kT/\chi\right)^{1/2}\;
\sum_{n=0}^5 a_n\left[\log(kT/\chi)\right]^n \;,
\end{equation}
when $\chi/10 \le kT \le 10\chi$, and
\begin{equation}
C_{1f} = n_e (kT/\chi)^{-1/2}\;\left[\gamma \ln(kT/\chi)+
\sum_{n=0}^2 \beta_n (\chi/kT)^n\right]\;,
\end{equation}
when $kT>10\chi$.  The values of the coefficients are given in Tables 1 and 2.
\medskip
\medskip
\centerline{Table 1}
\medskip
\centerline{
\begin{tabular}{|l|l|l|l|l|l|l|}
\hline
       & $a_0$ & $a_1$ & $a_2$ & $a_3$ & $a_4$ & $a_5$ \\ 
       \hline
HI&2.3742E-08&-3.6866E-09&-1.0366E-08&-3.8010E-09&3.4159E-09&1.6834E-09  \\ \hline
HeII&3.4356E-09&-1.6865E-09&-6.9236E-10&9.7863E-11&1.5591E-10&6.2236E-11  \\ \hline
\end{tabular}
}
\vskip 0.3 in
\centerline{Table 2}
\medskip
\centerline{
\begin{tabular}{|l|l|l|l|l|}
\hline
       & $\gamma$ & $\beta_0$ & $\beta_1$ & $\beta_2$  \\
       \hline
HI&2.4617E-08&9.5986E-08&-9.2463E-07&3.9973E-06  \\ \hline
HeII&3.0772E-09&1.1902E-08&-1.1514E-07&5.0489E-07  \\ \hline
\end{tabular}
}

\bigskip
\medskip

The radiative ionization rate from the ground state is given by the expression
\begin{equation}
R_{1f} = \int_\chi^\infty\sigma_{1f}(\epsilon) c n_\epsilon d\epsilon\;,
\end{equation}
where $n_\epsilon$ is the differential number
density of ionizing UV photons, normalized so that
\begin{equation}
\int_0^\infty \epsilon n_\epsilon d\epsilon = u_{uv}\;.
\end{equation}
The ionization cross-section for hydrogenic atoms (e.g., HI and HeII)
in their ground state is given by 
\begin{equation}
\sigma_{1f} (\epsilon)  = {8 h^3 \over 3^{3/2} \pi^2 e^2 m_e^2 c Z^2}
\left({\chi\over\epsilon}\right)^3\;\left[8\pi\; 3^{1/2}
\left(\chi\over\epsilon\right) {e^{-4(\alpha / \beta)
\cot^{-1} (\alpha / \beta)}
\over 1-e^{-2\pi (\alpha/ \beta)}}\right]\;,
\end{equation}
where $Z$ is the atomic number, $\alpha$ is the fine structure constant,
$\beta = v/c$ is the ratio of the free electron speed
to that of light, and the term in square brackets is an analytical
expression for the gaunt factor (Spitzer 1978).  

The total radiative recombination rate for hydrogenic atoms is 
\begin{equation}
\sum_i R_{fi} = n_e \sum_i <v \sigma_{fi} (E)>\;,
\end{equation}
where $v$ and $E$ are, respectively, the speed and kinetic 
energy of the electrons prior to capture, 
\begin{equation}
\sigma_{fi} (E) = {2^4\over 3^{3/2}}{he^2\over m_e^2 c^3}\;
{\chi^2\over (E+\chi/i^2) E} {g\over i^3}
\end{equation}
is the cross-section for radiative capture to the $i$-th 
state, and the brackets $<>$ denote an average over
the thermal electron distribution function.
The sum over $i$ may be truncated at $i = 20$ without introducing
any appreciable error.
For simplicity, the gaunt factors $g$ are set equal to unity,
therefore introducing a possible error that ranges from $\sim$ 3\% at $T = 10^4$
K to $\sim$ 20\% at temperatures in excess of $10^6$ K.  We
note, however, that our final results are not subject to the large
errors incurred at temperatures $\geq 10^6$ K.

Values of the rate coefficients as a function of the ambient
temperature $T$ are shown in Figure 2.
Note that the radiative recombination rates are
lower at all temperatures than the electron self-collision
rate, as discussed above.  For these conditions then,
the ionization equations are easily solved to 
determine the atomic number densities.  
The results of these calculations are shown in Figure 3, and these help to
justify the assumption stated earlier that the plasma is highly ionized for the
environment in the Sgr A East shell.

\section{The Positron Population}
In this section, we calculate the positron distribution, and
determine whether these particles thermalize before annihilating. We then consider
how to calculate the present positron number 
density for a given evolution of the Sgr A East environment.

\subsection{The Positron Annihilation Rates}
Positron annihilation in a hydrogen and helium plasma can occur via three
dominant mechanisms: direct (in-flight) annihilation; radiative capture; and charge
exchange. The first of these processes, in which a free positron
annihilates with an ambient electron without first forming positronium,
results in the production of two photons.  The direct annihilation
rate for a positron of kinetic energy $E_+$ interacting with an isotropic
Maxwell-Boltzman distribution $n_e f_{mb}(E_-)$ of electrons
(where $E_-$ is the electron kinetic energy) is given by the expression
\begin{equation}
R_{da} (E_+) = c n_e\;\int\;f_{mb}(E_-) 
dE_-\;\int_{-1}^1\;{d\mu\over 2} f_{kin} (E_+, E_-)
\;\sigma_{da} (E_+, E_-)\;,
\end{equation}
where 
\begin{equation}
f_{kin} = \left[\beta_-^2 + \beta_+^2 - \beta_-^2\beta_+^2
\;(1-\mu^2) - 2\beta_-\beta_+\mu\right]^{1/2}
\end{equation}
accounts for the kinematics of the two-particle motion, and
\begin{equation}
\beta_\pm = \left[{2E_\pm m_e c^2 + E_\pm^2\over (mc^2)^2 +
2E_\pm m_e c^2 + E_\pm^2}\right]^{1/2}
\end{equation}
are the positron/electron speeds in units of $c$.
The cross-section as a function of the relative Lorentz factor
$\gamma = (1-\beta^2)^{-1}$ can be determined using the 
plane-wave approximation and is found to be
\begin{equation}
\sigma_{da} = \left[2\pi\left({\alpha\over \beta}\right)\left(
1-e^{-2\pi\alpha/\beta}\right)^{-1}\right]\left[{\pi r_0^2\over\gamma+1}
\left({\gamma^2+4\gamma+1\over \gamma^2 - 1}\ln[\gamma+(\gamma^2-1)^{1/2}]
-{\gamma+3\over (\gamma^2-1)^{1/2}}\right)\right]\;,
\end{equation}
where
\begin{equation}
\beta = \left[{(1-\beta_+\beta_-\mu)^2 - (1-\beta_+^2)(1-\beta_-^2)
\over (1-\beta_+\beta_-\mu)^2}\right]^{1/2}
\end{equation}
is the relative speed (in units of $c$) between the electrons and positrons
(Coppi \& Blandford 1990), $r_0$ is the classical electron radius, and
the term in the first set of square brackets takes into account the
Coulomb attraction. 

Radiative capture produces positronium in either a singlet spin-0 state
(0.25 probability) or a triplet spin-1 state (0.75 probability).  
Annihilation from the singlet state produces two line photons 
whereas annihilation from the triplet state produces three 
continuum photons. The overall capture rate is given by summing over
the radiative capture rates into each state $i$:
\begin{equation}
R_{rc} (E^+) = \sum_i c\; n_e\; \int f_{mb}(E_-) 
dE_- \;\int_{-1}^{1} {d\mu\over 2}
f_{kin}(E_+, E_-)\; \sigma_{rc}(E_+, E_-)\;,
\end{equation}
where the cross-section is easily generalized from Equation (11) by
replacing $m_e$ with the reduced mass $m_e / 2$ and noting that the 
ionization energy from the ground state of positronium is $\chi = 6.8$ eV.  

The presence of HeII implies that charge exhange 
($e^+$ + HeII $\rightarrow$ Ps + HeIII) is an important 
mechanism by which positronium is formed. 
The rate at which this process occurs is given by
\begin{equation}
R_{ce} (E_+) = n_{HeII}\;\beta_+ c \sigma_{ce}(E_+)\;. 
\end{equation}
To our knowledge, no generally accepted theoretical or experimental
value exists for the cross-section of this process.
However, based on measurements of the cross-sections for
the $e^+$ + HI $\rightarrow$ Ps + HII  (Sperber et al. 1992) and
the $e^+$ + HeI $\rightarrow$ Ps + HeII reactions (Overton et al. 1993),
it is clear that the desired cross-section is peaked at 54.4 eV
(the ground-state ionization energy for HeII) with a value there
of $\sim \pi a_0^2$.  Assuming an energy profile similar to that
of the $e^+$ + HeI $\rightarrow$ Ps + HeII cross-section, we may therefore
adopt an empirically based cross-section of the form
\begin{equation}
\sigma_{ce}(E_+) = 2.4\times 10^{-16}\;\cm^2 \;
\left({E_+\over E_1}  - 1\right)\;e^{\left(1-{E_+ \over E_1}\right)}\;,
\end{equation}
for positron kinetic energies $E_+ > E_1 = 28$ eV.

The annihilation rates are shown as functions of the positron kinetic
energy in Figures 4-6 for three different values of the ambient
temperature.  The dotted lines, which represent the Maxwell-Boltzman
functions $f_{mb} (E_-)$ at each specified temperature, will be
introduced in \S\ 4.

\subsection{The Positron Cooling Rates}
We present here the cooling rates for a positron moving
through Sgr A East with a
Lorentz factor $\gamma_+ = [1-\beta_+^2]^{-1}$.
The dominant energy loss mechanisms are:
(1) Synchrotron losses at a rate 
\begin{equation}
{dE_s\over dt} = -{4\over 3} {\sigma_T c\over 8\pi}
\beta_+^2\gamma_+^2 B^2\;;
\end{equation}
(2) Compton scattering losses at a rate
\begin{equation}
{dE_\gamma\over dt} = -{4\over 3} m_e c^3\beta_+^2\gamma_+^2 \left[{u_{uv}
\sigma_c \left(x_{uv}\right)
 \over m_e c^2 +\gamma_+ kT_{uv}} +
{u_{ir}
\sigma_c \left(x_{ir}\right)
\over m_e c^2 +\gamma_+ kT_{ir}}\right]\;,
\end{equation}
where $x_{uv} = 2.7\gamma_+ kT_{uv} /m_e c^2$, $x_{ir} = 2.7\gamma_+
kT_{ir} /m_e c^2$, and
\begin{equation}
\sigma_c (x) = {3\over 4} \sigma_T \left[{1+x \over x^3}\left(
{2x (1+x)\over 1+2x} - \ln (1+2x)\right)
+{1\over 2x} \ln (1+2x) - {1+3x \over (1+2x)^2}\right]\;;
\end{equation}
(3) Bremmstrahlung emission losses at a rate 
\begin{equation}
{dE_B\over dt} = -4\alpha r_o^2 m_e c^3 (2 n_H + 6 n_{He}) 
\ln (2\gamma_+ - 1/3)\gamma_+\;;
\end{equation}
and (4) Coulomb losses at a rate of
\begin{equation}
{dE_C\over dt} = -m_e c^3 n_e \int f_{mb} (E^-) dE^- \int_{-1}^{1} 
{d\mu \over 2} \; f_{kin} <\sigma \Delta \gamma>\;,
\end{equation}
where 
\begin{equation}
<\sigma \Delta \gamma> = 
{3\sigma_T (\gamma_- - \gamma_+) \over 64 \epsilon^2} \\
\left[\left(4 \ln \Lambda + 4 \ln 2\right)
\left({\epsilon^2 + p^2 \over p^2}\right)^2
-2\left({8\epsilon^4 - 1 \over p^2 \epsilon^2}\right)
+{12\epsilon^4+ 1 \over \epsilon^4} -{8p^2\over 3}
{\left(\epsilon^2 + p^2\right)\over \epsilon^4}
+{2 p^4\over\epsilon^4}\right]
\end{equation}
is given in terms of the parameters
$\epsilon^2 = (1+\gamma)/2$ and $p^2 = \epsilon^2 - 1$, 
where $\gamma$ is the relative Lorentz factor (Coppi and Blandford 1990).
The corresponding cooling rates,
defined by the general expression $R = E_+^{-1} (dE/dt)$, 
are plotted in Figure 7 for an ambient temperature of $10^5$ K.
We note that only the Coulomb cooling rate depends on the
temperature, with the low-energy turnover occuring at an
energy of $\sim kT$.

\subsection{The Fate of Positrons in Sgr A East}
It is clear from the results shown in Figures 4-7 that
the total positron cooling rate is much greater than
the total positron annihilation rate over the entire
energy range.  In addition, for energies below $\sim 
10^8$ eV, the total cooling rate increases with 
decreasing particle energy.  Taken together, these results 
imply that particles produced in Sgr A East with energies
$\leq 10^8$ eV thermalize to the same temperature
as the ambient electrons on a time scale
equal to the inverse of their initial cooling rate, 
and do so before they annihilate.  These results are
summarized in Figure 8, which shows the total annihilation (short-dash) and
cooling (long-dash) rates assuming an ambient temperature
of $T = 10^5$ K
along with the injection function (solid line, right-hand scale)
for the Sgr A East positrons.  The two horizontal dotted lines
mark the inverse of $10^4$ and $10^5$ years, and the energy at 
which they first intersect the injection function
then yields the threshold energy below which particles (i.e., those
to the left of the corresponding vertical lines)
have had sufficient time to thermalize within that time scale. 
A plot of the threshold energy $E_t$ as a function of
time ranging between $10^4$ and $10^5$ years is shown in Figure 9.
Figure 10 then illustrates the rate per unit volume with which positrons
thermalize in Sgr A East (found by integrating the injection 
function up to the threshold energy) 
as a function of its age.  This value falls well
short of the total rate per unit volume with which positrons
are produced in Sgr A East ($2.7\times 10^{-18}$ cm$^{-3}$ s$^{-1}$)
since $E_t$ is smaller than the energy at the peak of the injection 
function.

In other words, Sgr A East is apparently producing positrons at 
a rate that is currently much higher than that with which they cool and thermalize.
The positrons in this remnant are being ``stockpiled.''

\section{Evolutionary Models of Sgr A East}
The number density of {\it thermal} positrons populating Sgr A East
as a function of its age $t^*$
is found by integrating the evolution equation
\begin{equation}
{dn_+\over dt} = \left[\int_0^{E_t (t^*)} I_+ (E) dE\right] -
\left[n_+ \left(<R_{da}> + <R_{ce}> + <R_{rc}>\right)\right]\;,
\end{equation}
where $I_+$ is the injection function
shown in Figure 1 and the terms in the $<>$ brackets 
represent the annihilation rates 
(da - direct annihilation; ce - charge exchange; rc - 
radiative capture) averaged over the thermal positron
distribution function.  Note that the term in the first
square brackets represents the rate at which positrons 
are thermalized in Sgr A East.

We first consider a static evolutionary model for Sgr A East
where both the temperature and thermalization rate are constant 
in time.  For this case, the number density of positrons increases
with age until the annihilation rate matches the thermalization
rate.  Assuming that all thermal 
positrons form positronium before annihilating, each positron
produces an average of $0.25\times 2 + 0.75 \times 3 = 2.75$
annihilation photons.  A reasonably good estimate for the ensuing
flux at Earth, found by multiplying the product of
the injection rate (which is clearly larger than the
thermalization/annihilation rate)
and the Sgr A East volume by $2.75$ to  find
the total rate of annihilation radiation photons produced
in Sgr A East, and then dividing by $4\pi D^2$ (where $D = 8.5$ 
kpc is the distance to the Galactic Center), yields a value of
$6.5\times 10^{-6} \cm^{-2}\;\sec^{-1}$.  This number 
is clearly much smaller than the measured flux arriving at earth
from the bulge.

Is it possible then that a time-dependent positron production rate
in Sgr A East could account for the current high rate of annihilation?
Let us consider the possibility
that positrons were stockpiled in Sgr A East during its infancy,
and are now being depleted
at a rate exceeding the current thermalization rate.  This feat
can be accomplished in two ways.  First, the positron injection rate in Sgr
A East may have been much higher earlier in its life.
Indeed, the energy content of the relativistic
proton population for Case 1 of Melia et al. (1989)
is $7.7\times 10^{51}$ ergs, or roughly
10\% of the energy associated with Sgr A East.  (We note, however,
that this value is really an upper limit, as it was derived
assuming steady state conditions.)  In that case, the 
number density of positrons populating Sgr A East would now
be $\sim 10^{-2}$ cm$^{-3}$.  In contrast, for the static
evolutionary model discussed above,
the number density of positrons injected into Sgr A East
during the assumed lifetime of 75,000 years is $\sim 6\times
10^{-6}$ cm$^{-3}$.  This difference suggests that the initial injection
rate may have been as high as $10^4$ times greater than
what is observed today.  

We therefore consider two cases
for the evolution of the injection rate, both of the form
\begin{equation}
I_+ (t) = A\; I_+ (t^*) \exp(-t/\tau)\;,
\end{equation}
where $A = 10^2$ for Case A and $A = 10^4$ for Case B.  The 
value for $\tau$ is then well-defined for the assumed age 
$t^*$ of 75,000 years.

Another reason why positrons may have been stockpiled in Sgr A East is 
that its temperature decreases with age.  This can be seen by
considering the annihilation
rates averaged over the thermal positron distribution 
as a function of temperature.  As shown in Figure 11,
the averaged charge-exchange rate is strongly peaked at
$\sim 10^5$ K, where the overlap between the Maxwell-Boltzman distribution
function and the charge-exchange rate produces a clear maximum
(see Figures 4 - 6).  
It is intriguing that the lower limit of 0.94
for the positronium fraction observed from the Galactic Center
indicates that the temperature where the annihilation radiation is produced
must lie in the range between $3.2\times 10^4$ K and
$3.2\times 10^5$ K.  This result is clearly seen from the plot
of the positronium fraction 
\begin{equation}
f_{Ps} = {<R_{ce}>+<R_{rc}> \over <R_{ce}> + <R_{rc}>
+ <R_{da}>}\;
\end{equation}
versus temperature, as shown in Figure 12.
These results are somewhat consistent
with the observed line-width of $2.5$ keV, since the
expected relation
\begin{equation}
{\Delta \epsilon\over\epsilon} \approx 3.2\times 10^{-3}
\;\left({T\over 10^4 \kelvin}\right)^{0.41}\;,
\end{equation}
(Wallyn et al. 1996)
then implies a temperature of $\approx 2.8\times 10^4$ K.
We caution, however, that  the above line-width relation
is somewhat suspect (as noted by Wallyn et al.).
We therefore assume that the initial temperature of
Sgr A East may have been $10^7$ K and that it decays exponentially 
so that the temperature will be $10^5$K at an age of 
$10^5$ years.  

\section{Results and Discusson}

We have integrated Equation (26) for the two evolutionary cases (A and B)
discussed above.  
The number densites as a function of age for these situations
are shown in Figure 13 as solid lines.  The dotted lines represent the number
densities stockpiled in the event that the annihilation rates are zero.
To gain insight into these results, we plot the rate per unit volume 
at which positrons are thermalized (dotted line) and annihilated (solid and
dashed lines) as a function of age
for case B in Figure 14.  It is clear that the positrons are
indeed stockpiled when the thermalization rates are high,
and this stockpile is depleted when the charge-exchange
annihilation rate peaks.

The ensuing annihation spectrum is easily determined by noting that the
observed flux of line photons is given by
\begin{equation}
F_l = {N_+\over 4\pi D^2} \left(2 <R_{da}> + 0.5 <R_{rc}>
+ 0.5 <R_{ce}>\right)\;,
\end{equation}
and the observed flux of continuum photons is given by
\begin{equation}
F_c = {N_+ \over 4\pi\ D^2} \left(2.25 <R_{rc}> +
2.25 <R_{ce}>\right)\;,
\end{equation}
where $N_+$ is the total number of thermal positrons in Sgr A
East.  The total flux observed as a
function of age is presented in Figure 15
for the two cases discussed above.  As expected, 
the flux peaks at around $75,000$ years, where the
positrons are just beginning to be depleted.

It is clear by comparing the values in Figure
15 to the observed flux from the bulge that even ideal evolutionary
conditions cannot make up for the fact that only a small
fraction of the injected positrons will have thermalized and are
annihilating.  This point is emphasized by the requirement of
a very efficient cooling mechanism for the conditions in Sgr A East to account for the
observed annihilation radiation flux, as may be seen by
making the {\it ad hoc} assumption that
all of the injected positrons thermalize.  In that (extreme) case,
the resulting flux curve (as shown in Figure 16 for the case with 
$A = 10^2$) would agree 
well with the observations.

Our first conclusion is therefore that Sgr A East
is not (by itself) the source of annihilation radiation
from the Galactic Center, unless some efficient cooling
mechanism (not considered here) is thermalizing the positrons
on a time scale shorter than the age of this remnant.

However, since Sgr A East's age is much shorter than the age
of the Galaxy, it is reasonable to ask whether the process that
produced this relatively unique remnant was a singular event.  If not, then 
the accumulated ``stockpiles'' of positrons may
now be at the required level to account for the observed
annihilation rate.  Unfortunately, there is very little to
go by in terms of phenomenologically pinning down the rate
at which these explosive events have occurred.  But we can
at least use the properties of the bulge component of the
$e^+e^-$ annihilation radiation to infer what the minimum explosion
rate must have been.

Using the required positron density, and 
assuming that no special cooling rate, beyond what
we have considered here, is acting on the positrons, we estimate that
$\sim 30$ explosions (like the one that produced Sgr A East) must have
occurred over the past one million years if $A = 10^2$.  The number of
events drops as the value of $A$ increases, reaching about one event
per million years when $A = 10^4$.
This is interesting in view of
the fact that Sgr A East's age is estimated to be about 75,000 years,
which means that indeed we may be witnessing the most recent of
a series of catastrophic events that occur on average about once
every 30,000 to 1 million years.  
This lends indirect support to the idea (Khokhlov
and Melia 1996) that Sgr A East may have been produced by the tidal
disruption of a $\sim 2\;M_\odot$ star that ventured within $10$
Schwarzschild radii or so of the central black hole.  Given the 
stellar density in the vicinity of Sgr A*, such events are expected
to occur roughly once every 10,000 to 100,000 years.  The bulge
component of the $e^+e^-$ annihilation radiation may therefore be
the best evidence we have right now for the existence of 
accumulated relics of this ongoing stellar disruption process.

{\bf Acknowledgments} This research was partially supported 
by NASA under grants 
NAG5-8239 and NAG5-9205, and has made use of NASA's 
Astrophysics Data System Abstract Service.  FM is very grateful
to the University of Melbourne for its support (through a
Miegunyah Fellowship) and MF would like
to thank the John Hauck Foundation for partial support.

{}

%
%

\newpage

\begin{figure}[thb]\label{fig1.ps}
{\begin{turn}{0}
\epsscale{1.0}
\plotone{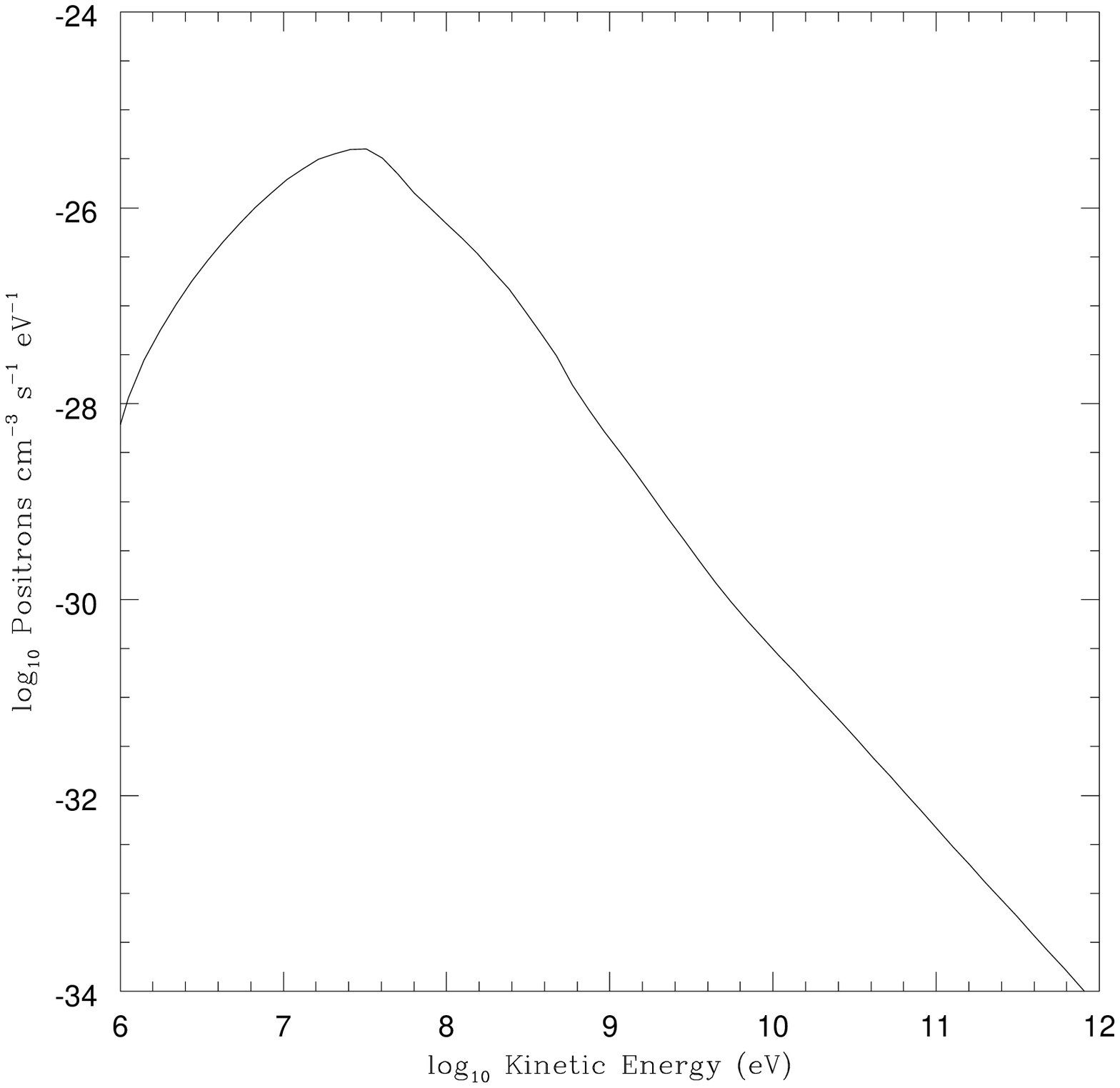}
\end{turn}}
\caption{The injection function $I_+ (E_+)$
for positrons produced via charged
pion decay in Sgr A East, assuming conditions adopted in Case 2 of Melia
et al. (1998).  The total rate per unit volume at which positrons
are produced, determined by integrating this function over
the positron kinetic energy, is equal to $2.7\times 10^{-18}$ 
cm$^{-3}$ s$^{-1}$.} 
\end{figure}

\clearpage
\begin{figure}[thb]\label{fig2.ps}
{\begin{turn}{0}
\epsscale{1.0}
\plotone{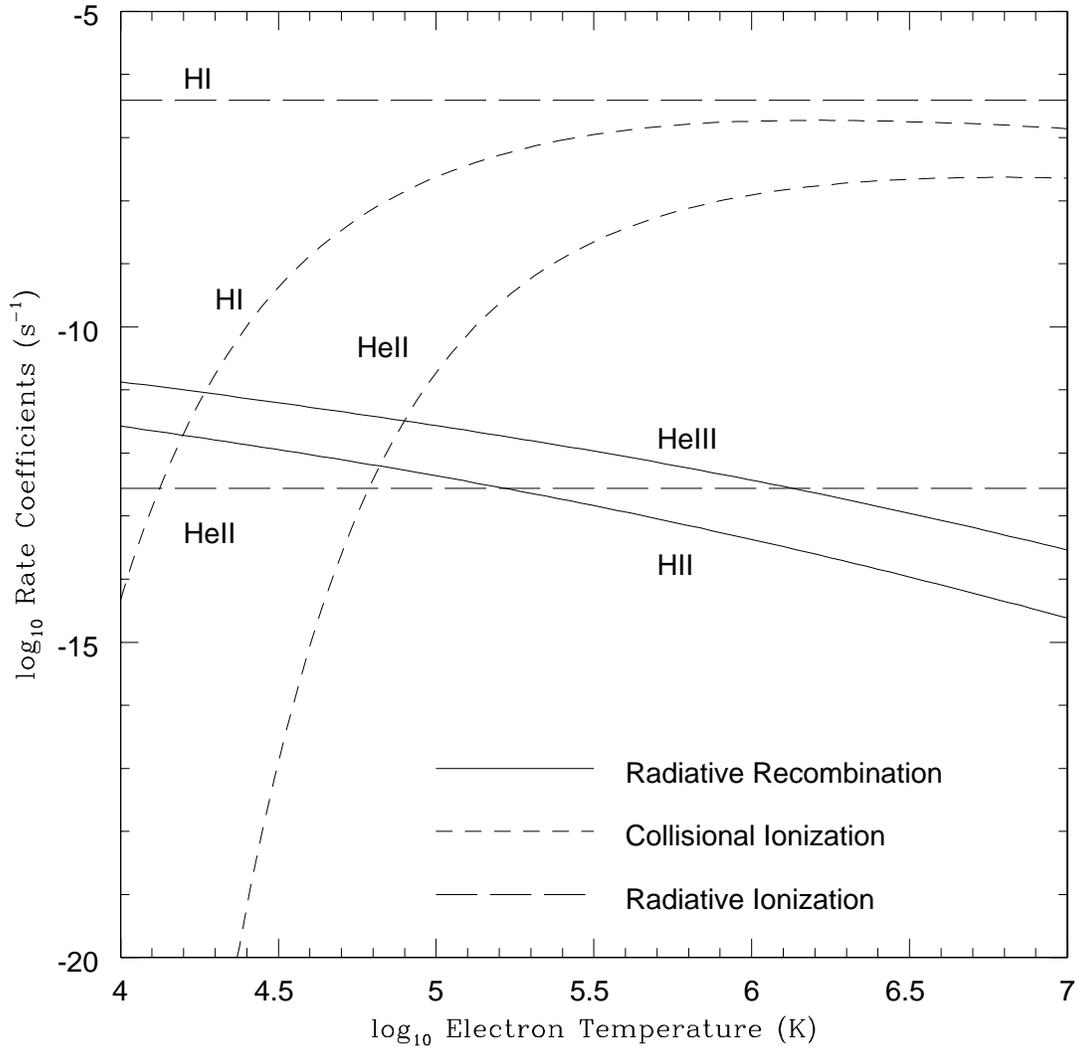}
\end{turn}}
\caption{Recombination and ionization rates for hydrogen and helium
in Sgr A East as a function of the ambient electron temperature. }
\end{figure}

\clearpage
\begin{figure}[thb]\label{fig3.ps}
{\begin{turn}{0}
\epsscale{1.0}
\plotone{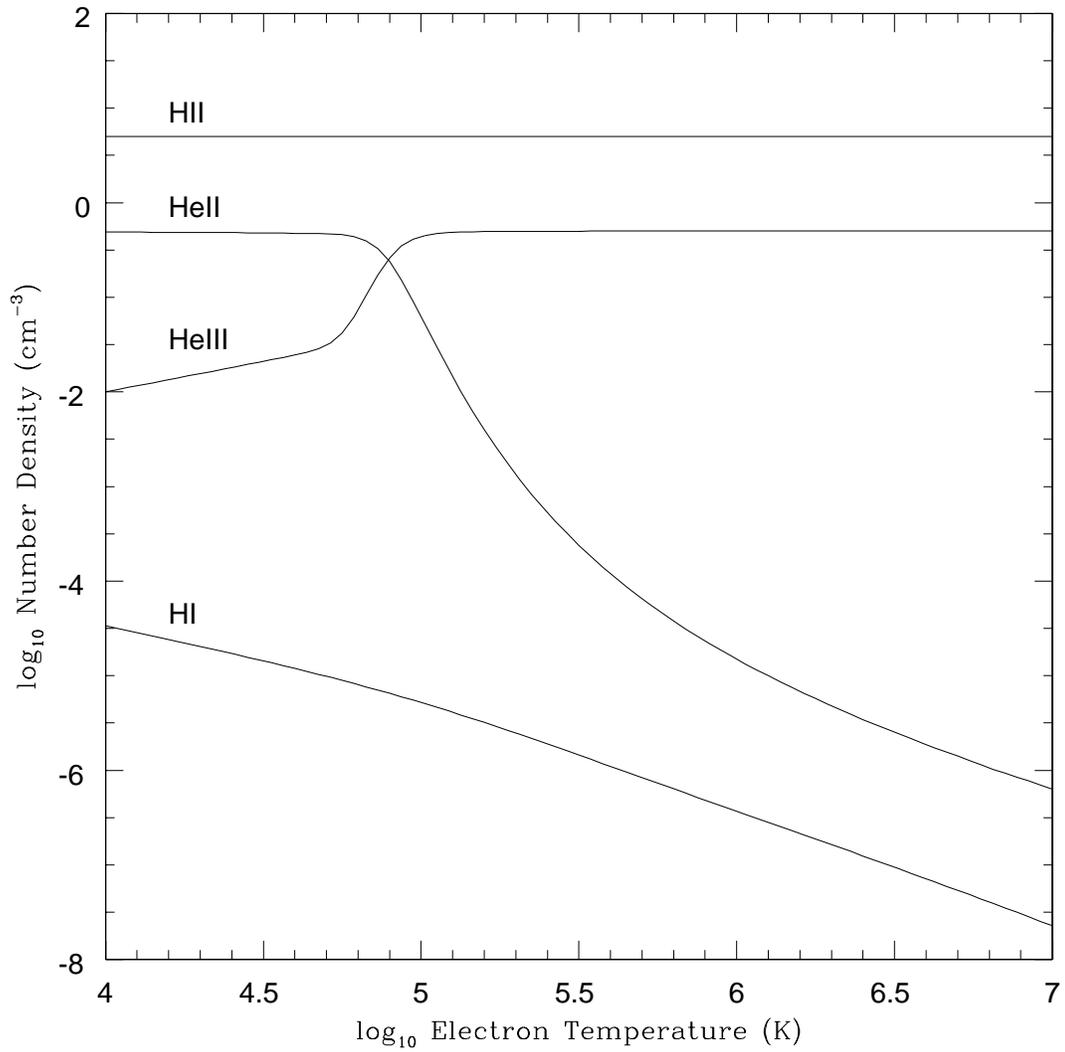}
\end{turn}}
\caption{The number density of neutral and ionized hydrogen and of singly and
doubly ionized helium in Sgr A East as a function of the ambient electron
temperature.  }
\end{figure}

\clearpage
\begin{figure}[thb]\label{fig4.ps}
{\begin{turn}{0}
\epsscale{1.0}
\plotone{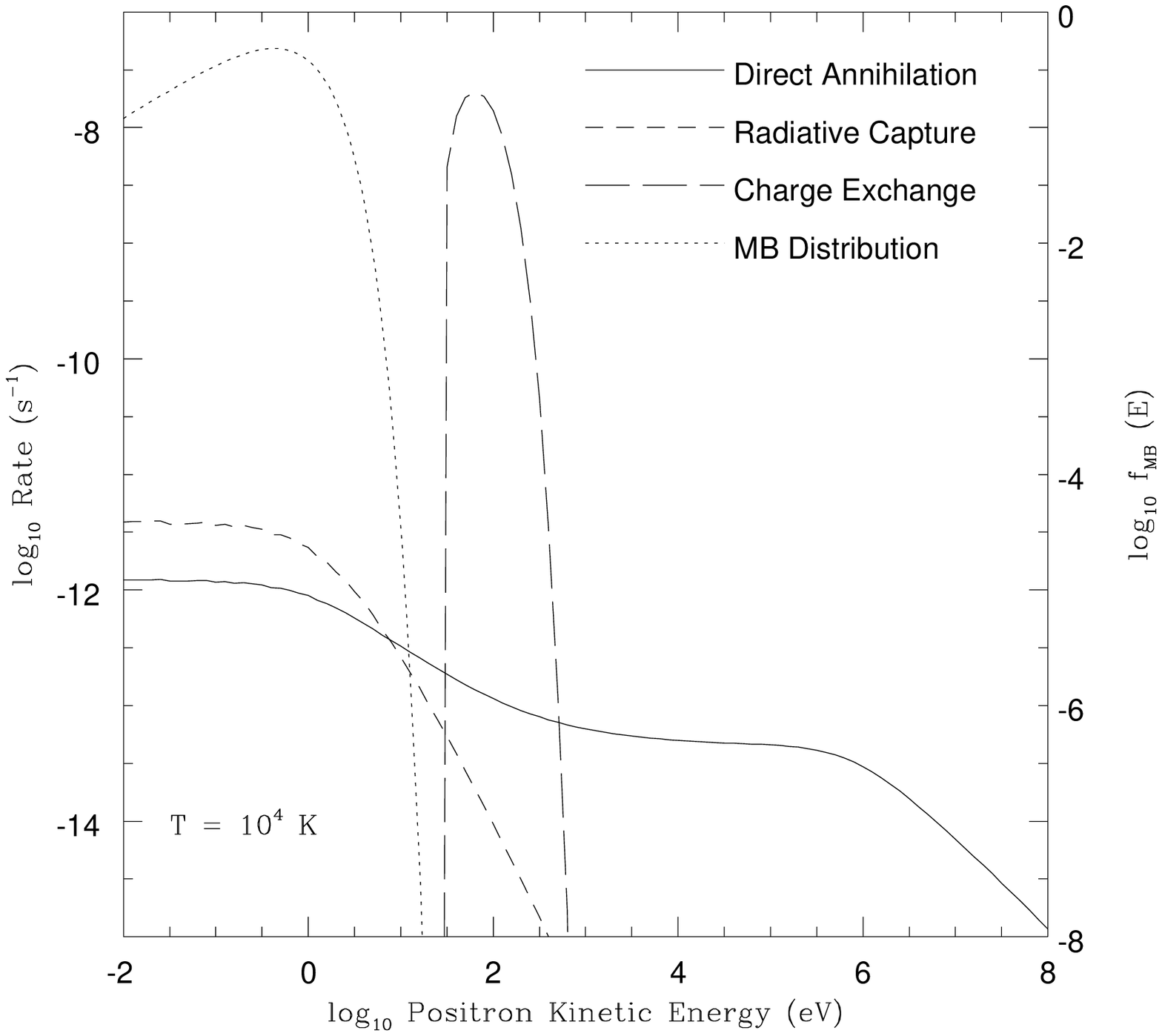}
\end{turn}}
\caption{Annihilation rates as a function of positron 
kinetic energy for an ambient temperature of $10^4$ K.}
\end{figure}

\clearpage
\begin{figure}[thb]\label{fig5.ps}
{\begin{turn}{0}
\epsscale{1.0}
\plotone{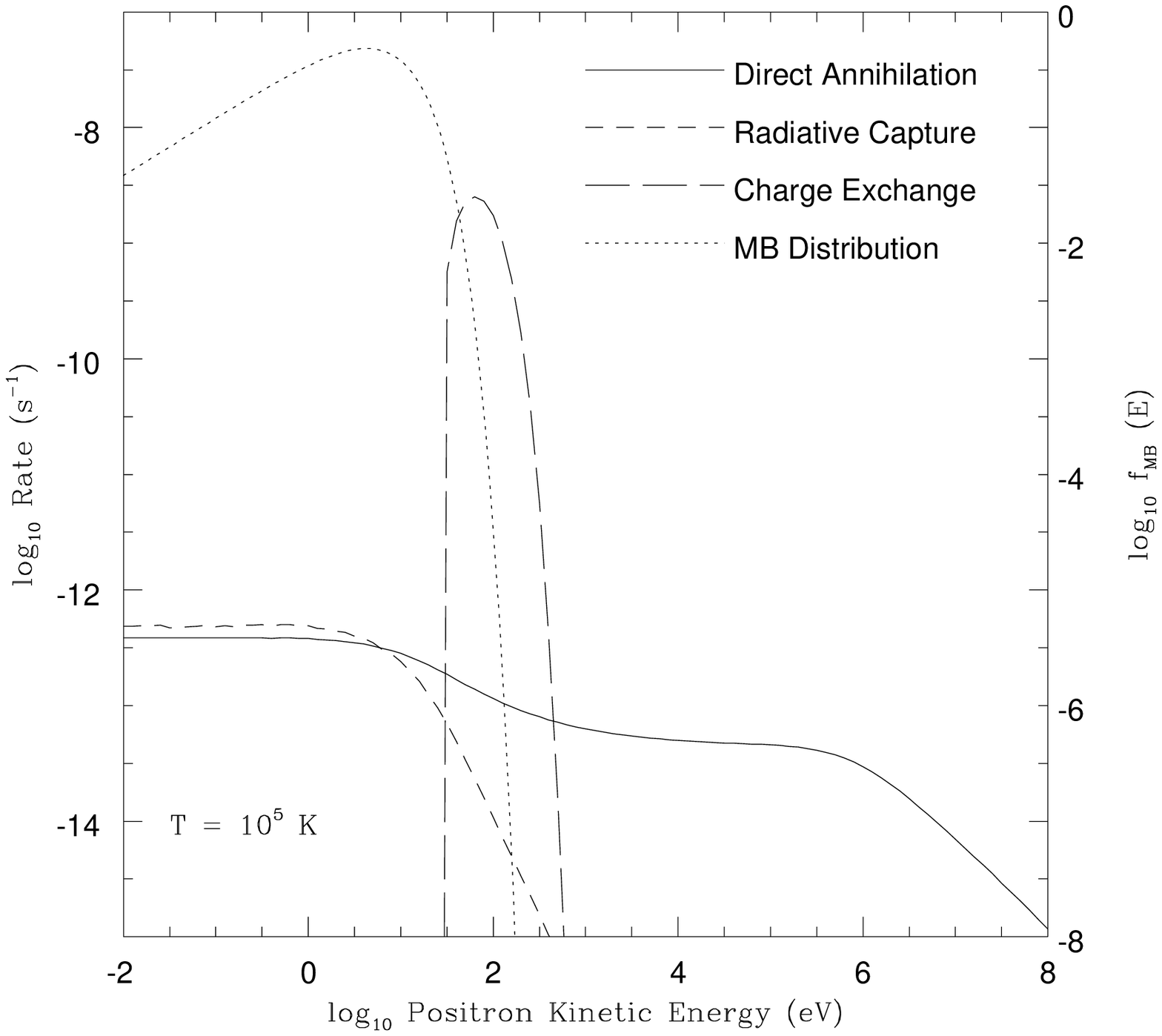}
\end{turn}}
\caption{Annihilation rates as a function of positron
kinetic energy for an ambient temperature of $10^5$ K.}
\end{figure}

\clearpage
\begin{figure}[thb]\label{fig6.ps}
{\begin{turn}{0}
\epsscale{1.0}
\plotone{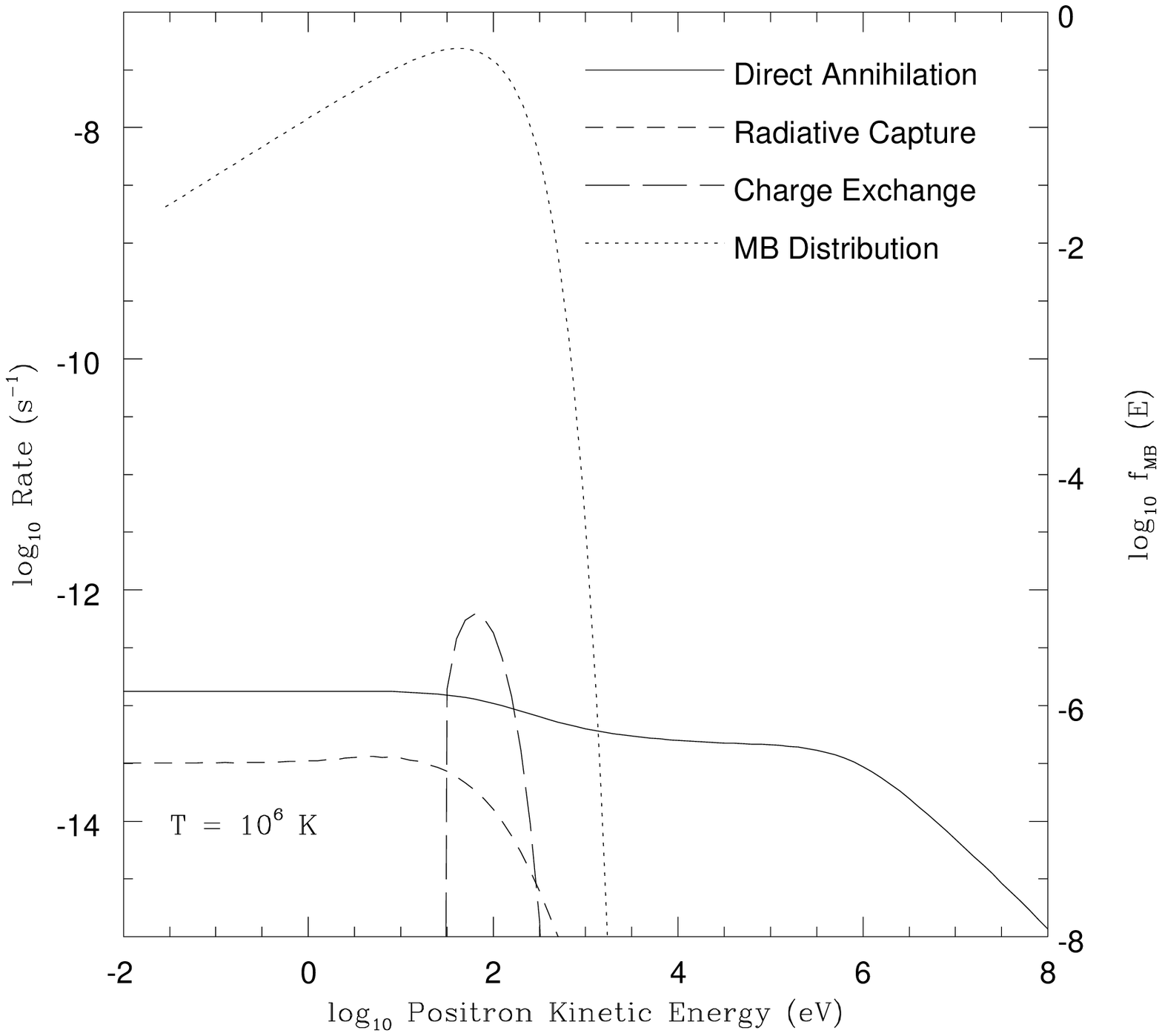}
\end{turn}}
\caption{Annihilation rates as a function of positron
kinetic energy for an ambient temperature of $10^6$ K.}
\end{figure}

\clearpage
\begin{figure}[thb]\label{fig7.ps}
{\begin{turn}{0}
\epsscale{1.0}
\plotone{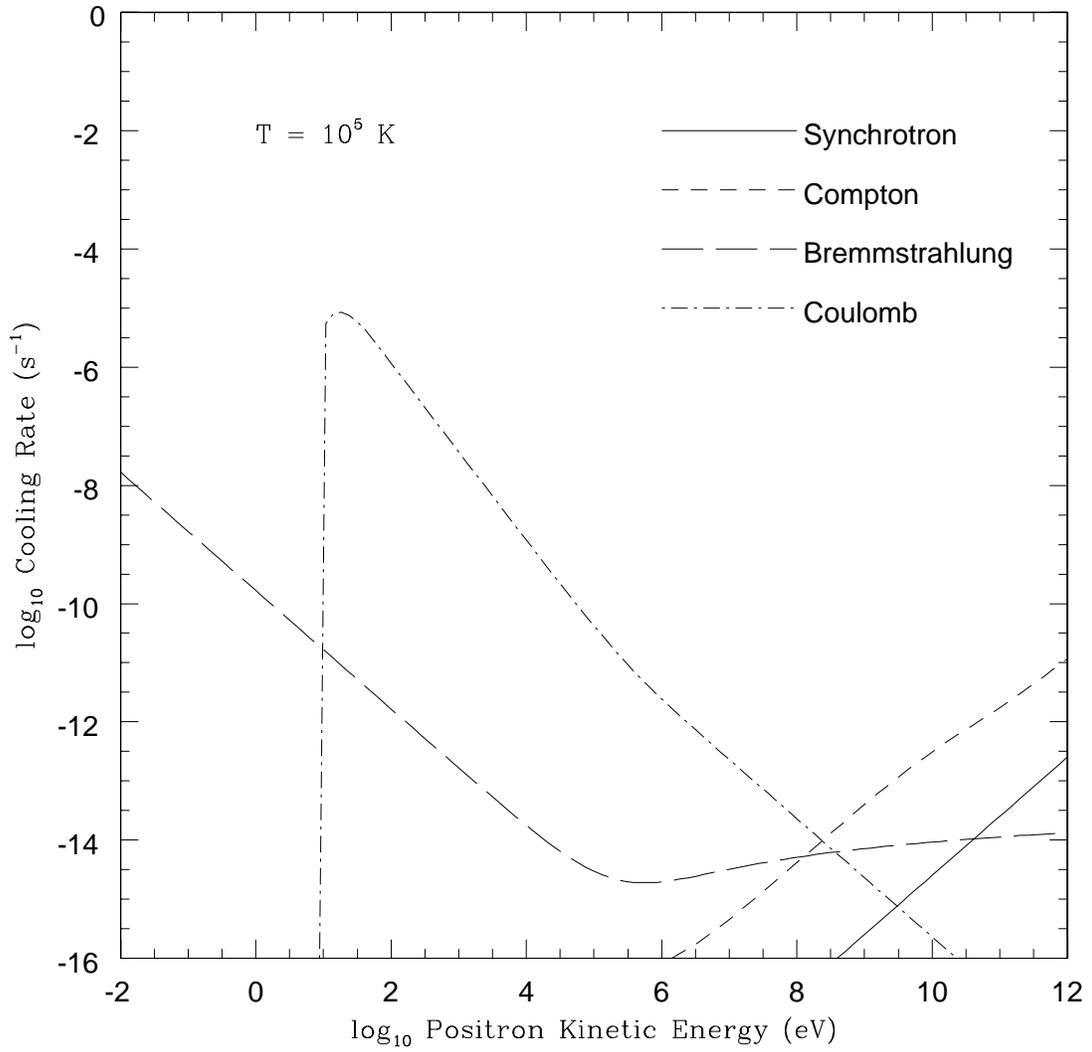}
\end{turn}}
\caption{Cooling rates as a function of positron
kinetic energy for an ambient temperature of $10^5$ K.}
\end{figure}

\clearpage
\begin{figure}[thb]\label{fig8.ps}
{\begin{turn}{0}
\epsscale{1.0}
\plotone{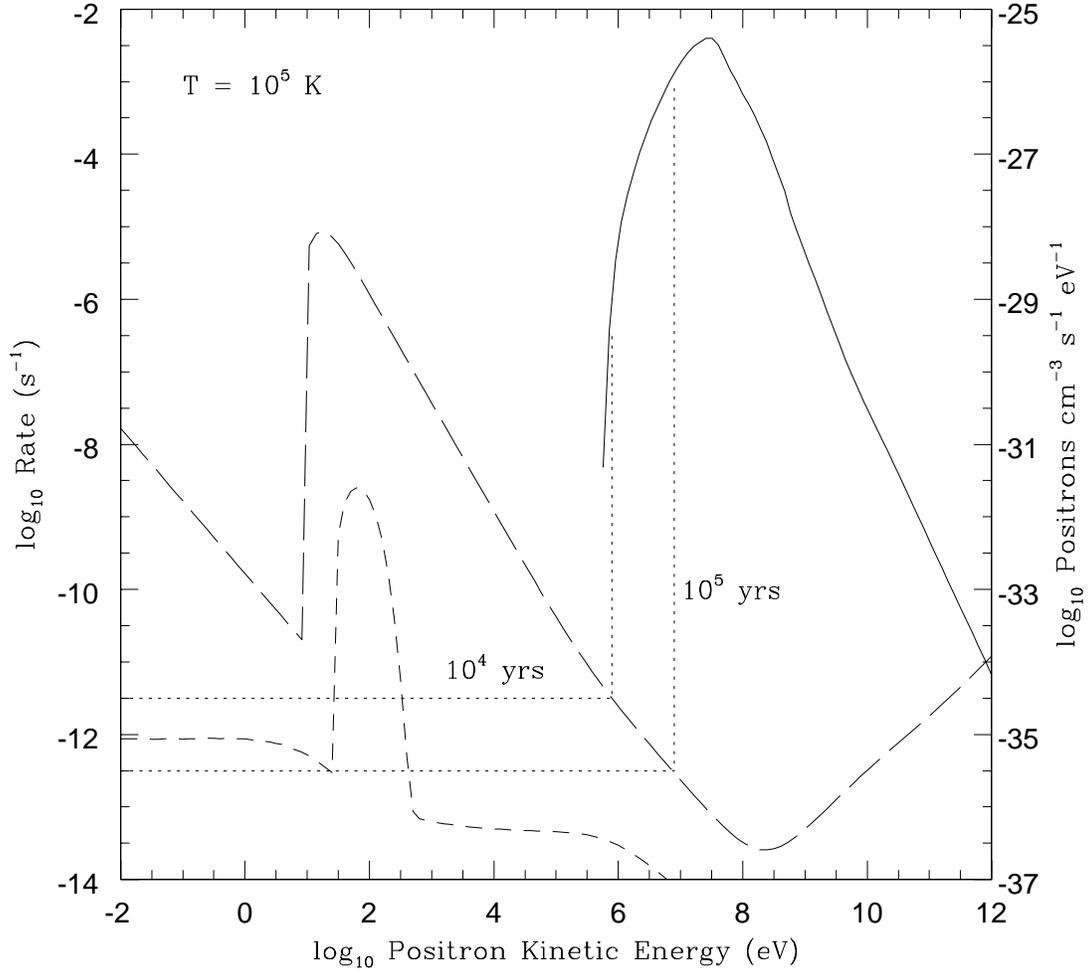}
\end{turn}}
\caption{A direct comparison between the total annihilation 
and cooling rates, and the injection function in terms of the positron
kinetic energy for an ambient temperature of $10^5$ K.}
\end{figure}

\clearpage
\begin{figure}[thb]\label{fig9.ps}
{\begin{turn}{0}
\epsscale{1.0}
\plotone{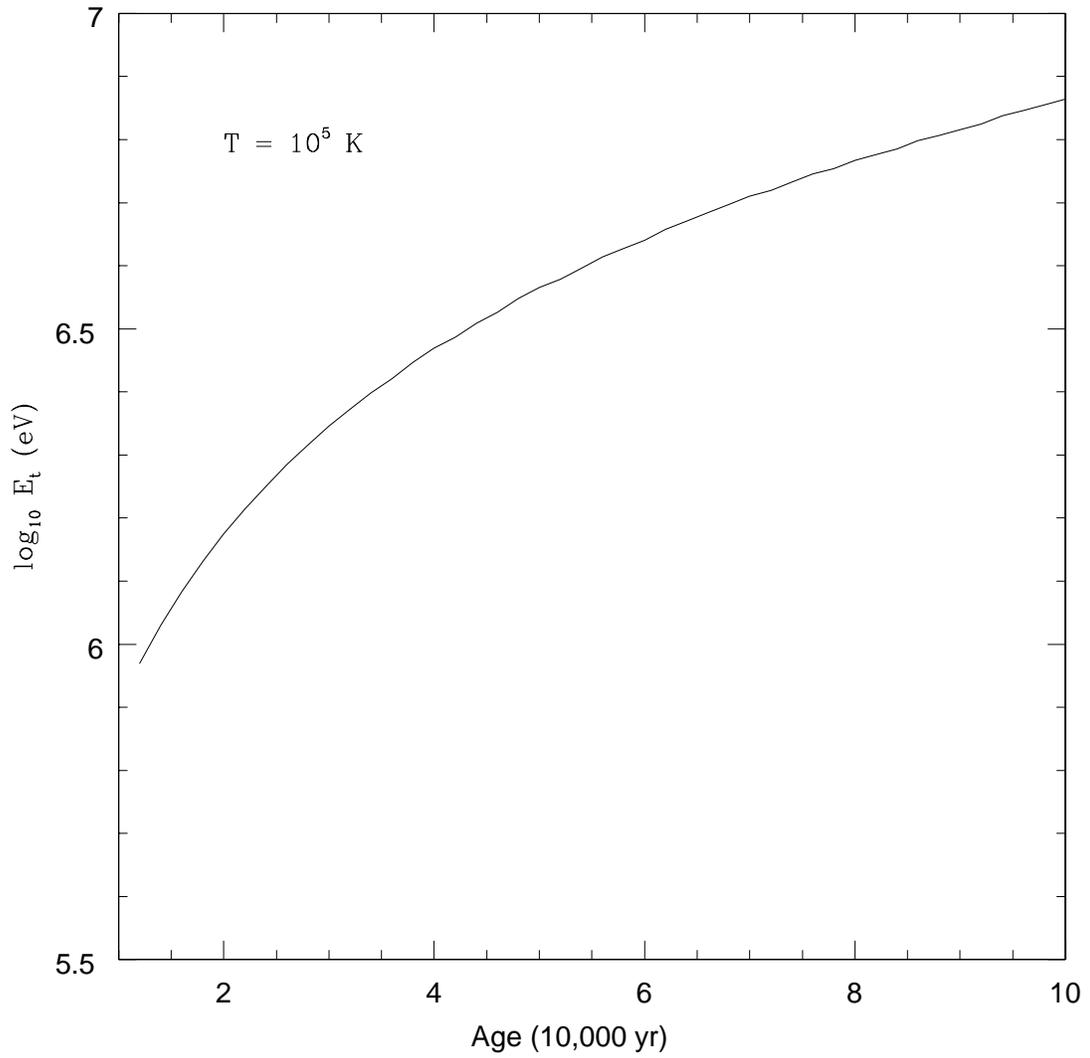}
\end{turn}}
\caption{The threshold energy below which positrons
thermalize on a time scale shorter than the age of Sgr A East
as a function of the remnant's age.}
\end{figure}

\clearpage
\begin{figure}[thb]\label{fig10.ps}
{\begin{turn}{0}
\epsscale{1.0}
\plotone{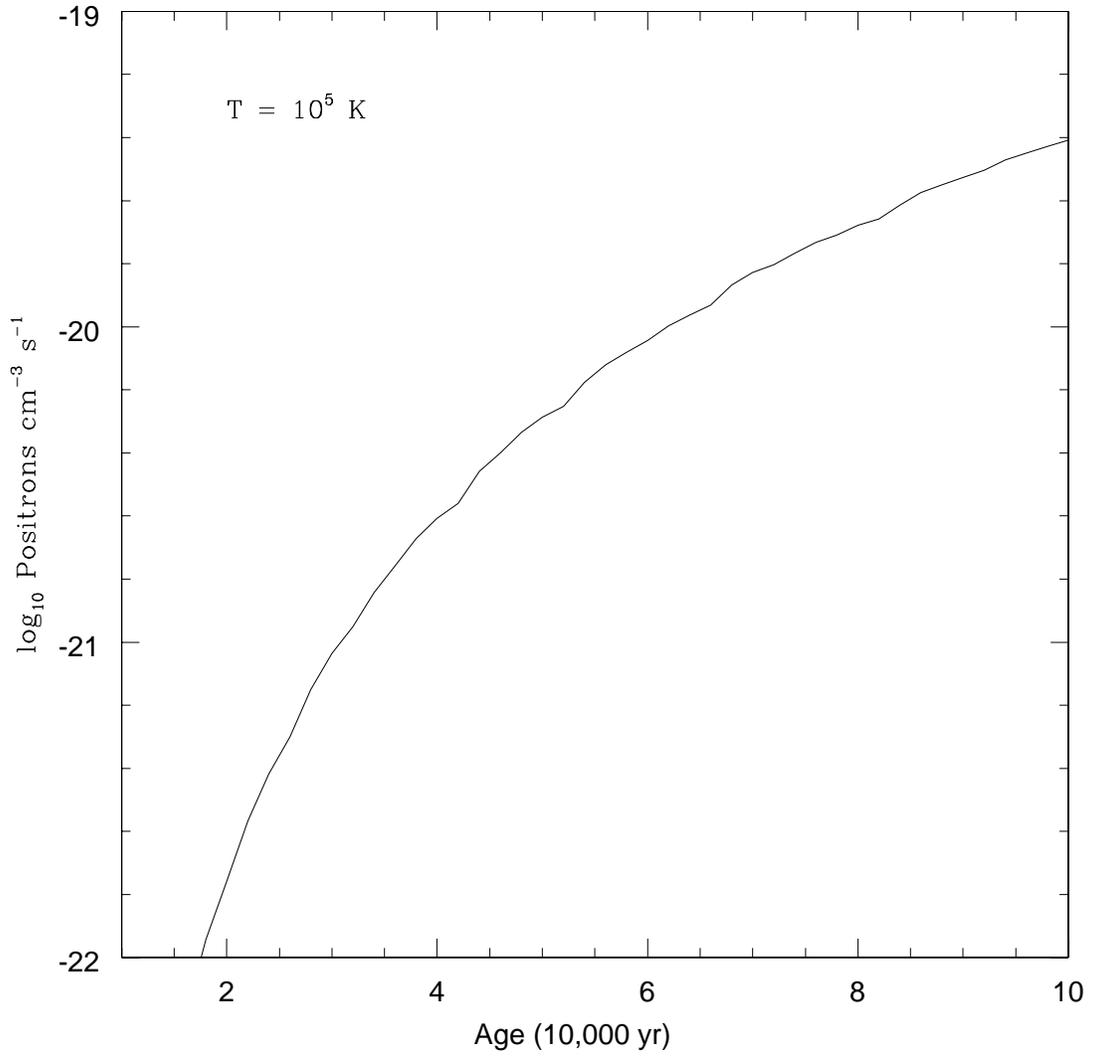}
\end{turn}}
\caption{The thermalization rate as a function of age for Sgr A East.}
\end{figure}

\begin{figure}[thb]\label{fig11.ps}
{\begin{turn}{0}
\epsscale{1.0}
\plotone{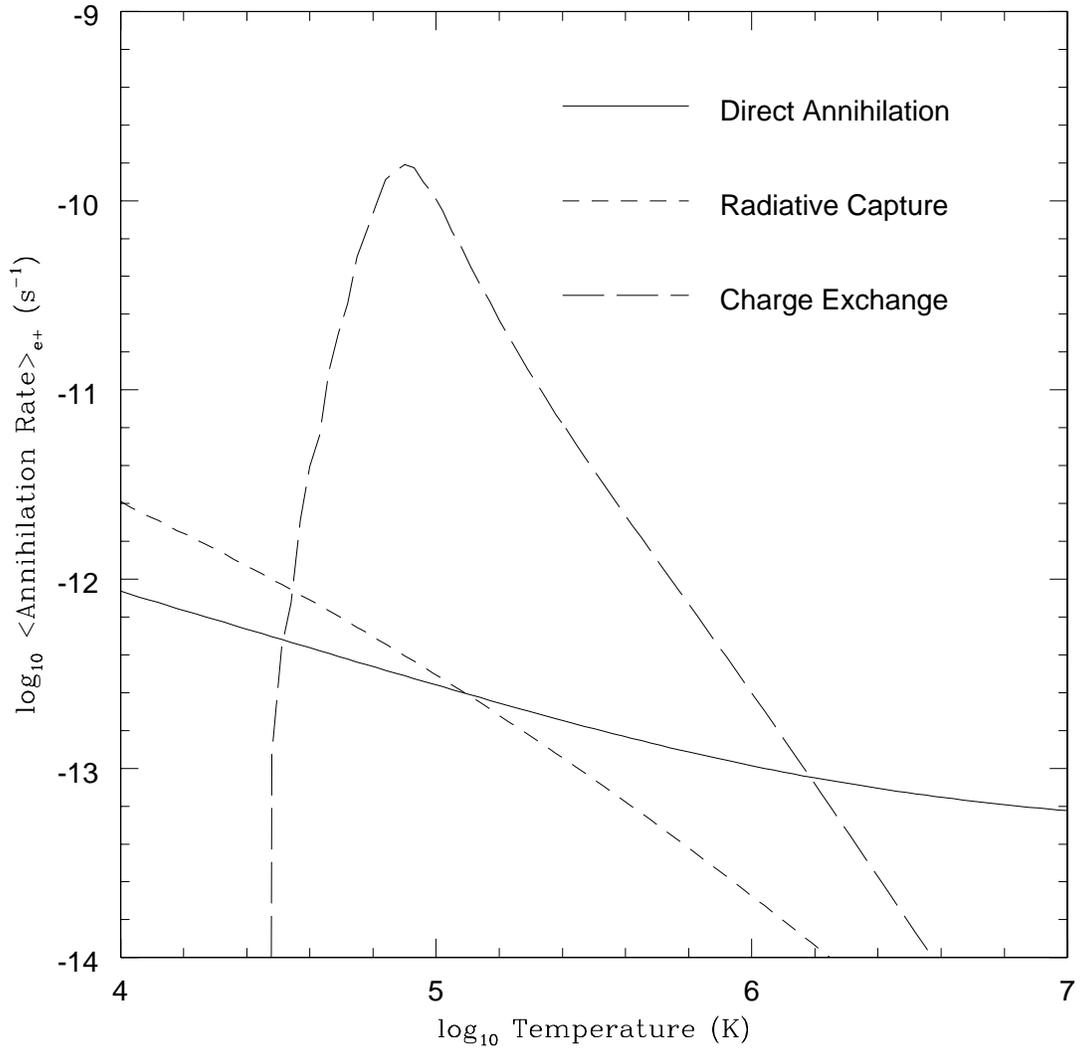}
\end{turn}}
\caption{The positron annihilation rates averaged over the positron 
distribution 
as a function of temperature.}
\end{figure}

\clearpage
\begin{figure}[thb]\label{fig12.ps}
{\begin{turn}{0}
\epsscale{1.0}
\plotone{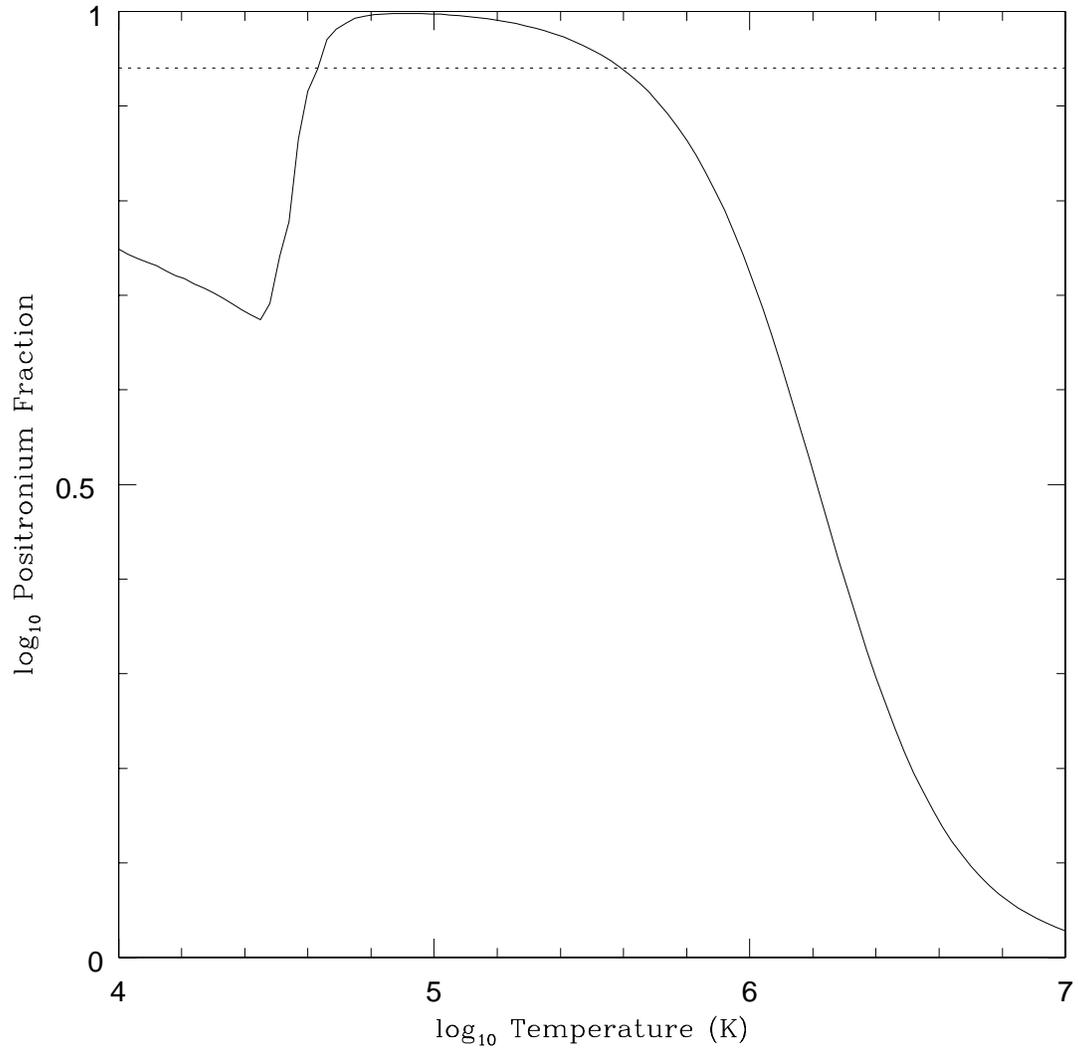}
\end{turn}}
\caption{The positronium fraction as a function of temperaure.}
\end{figure}

\clearpage
\begin{figure}[thb]\label{fig13.ps}
{\begin{turn}{0}
\epsscale{1.0}
\plotone{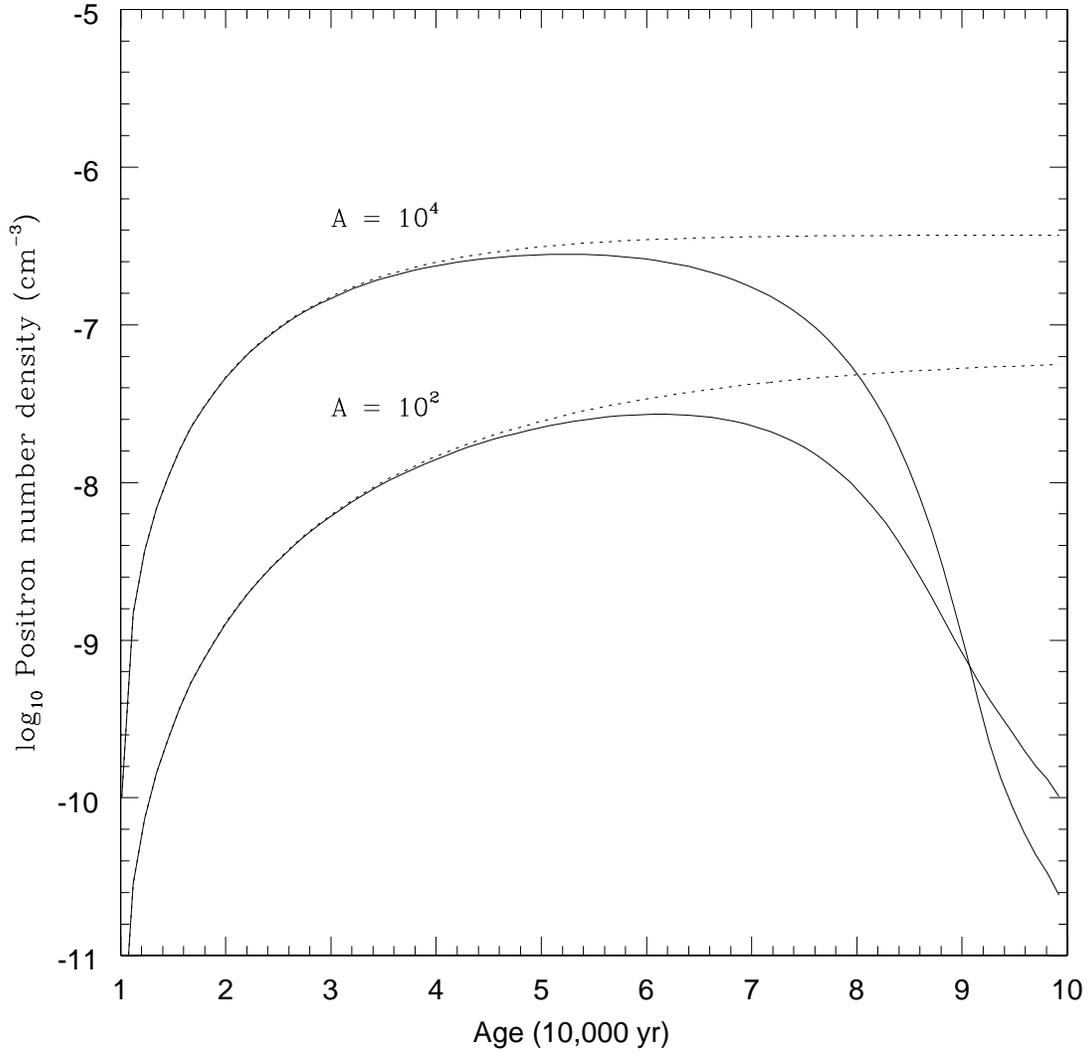}
\end{turn}}
\caption{The number density of thermal positrons for the two cases
considered.}
\end{figure}

\clearpage
\begin{figure}[thb]\label{fig14.ps}
{\begin{turn}{0}
\epsscale{1.0}
\plotone{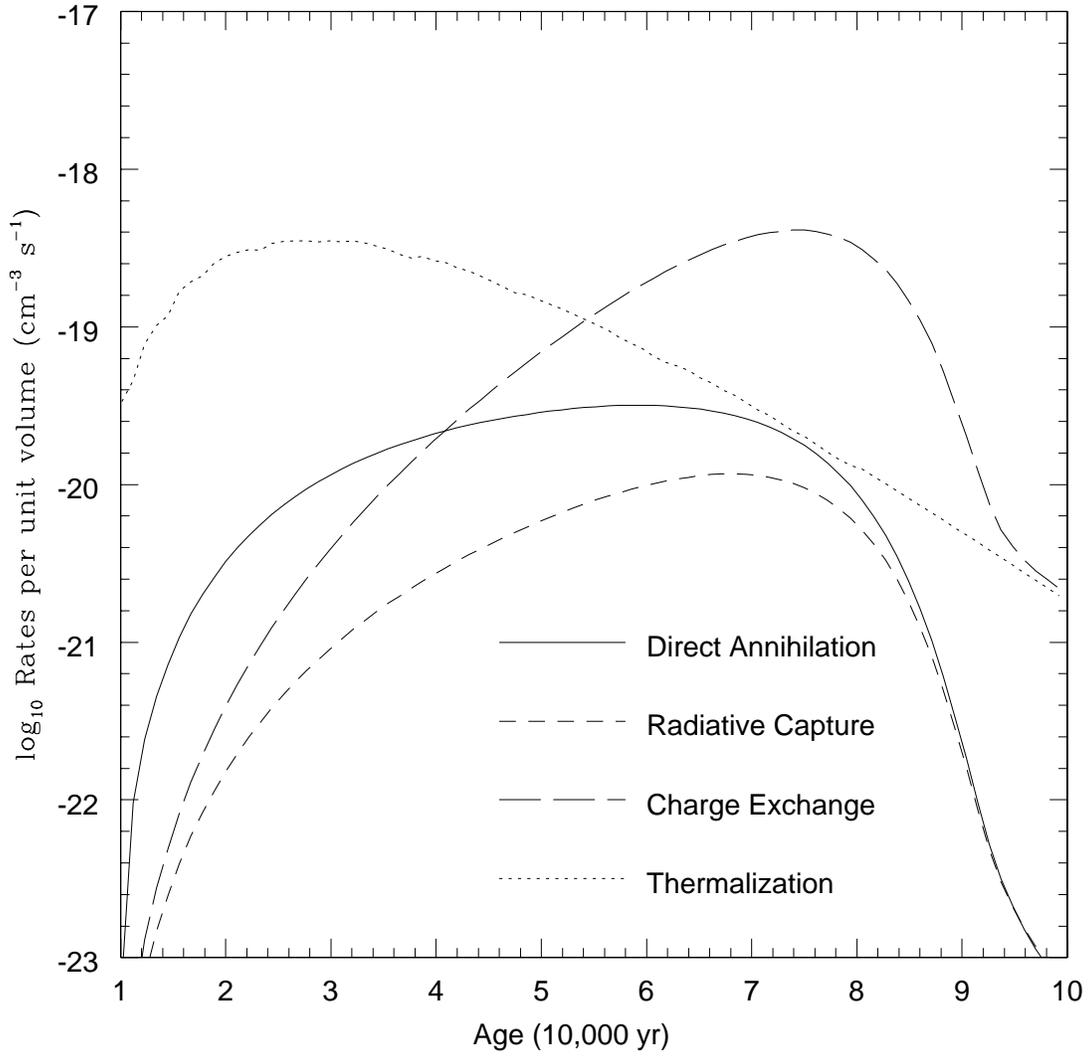}
\end{turn}}
\caption{A comparison between the rates per unit volume of thermalization and
annihilation as a function of age for $A=10^4$.}
\end{figure}

\clearpage
\begin{figure}[thb]\label{fig15.ps}
{\begin{turn}{0}
\epsscale{1.0}
\plotone{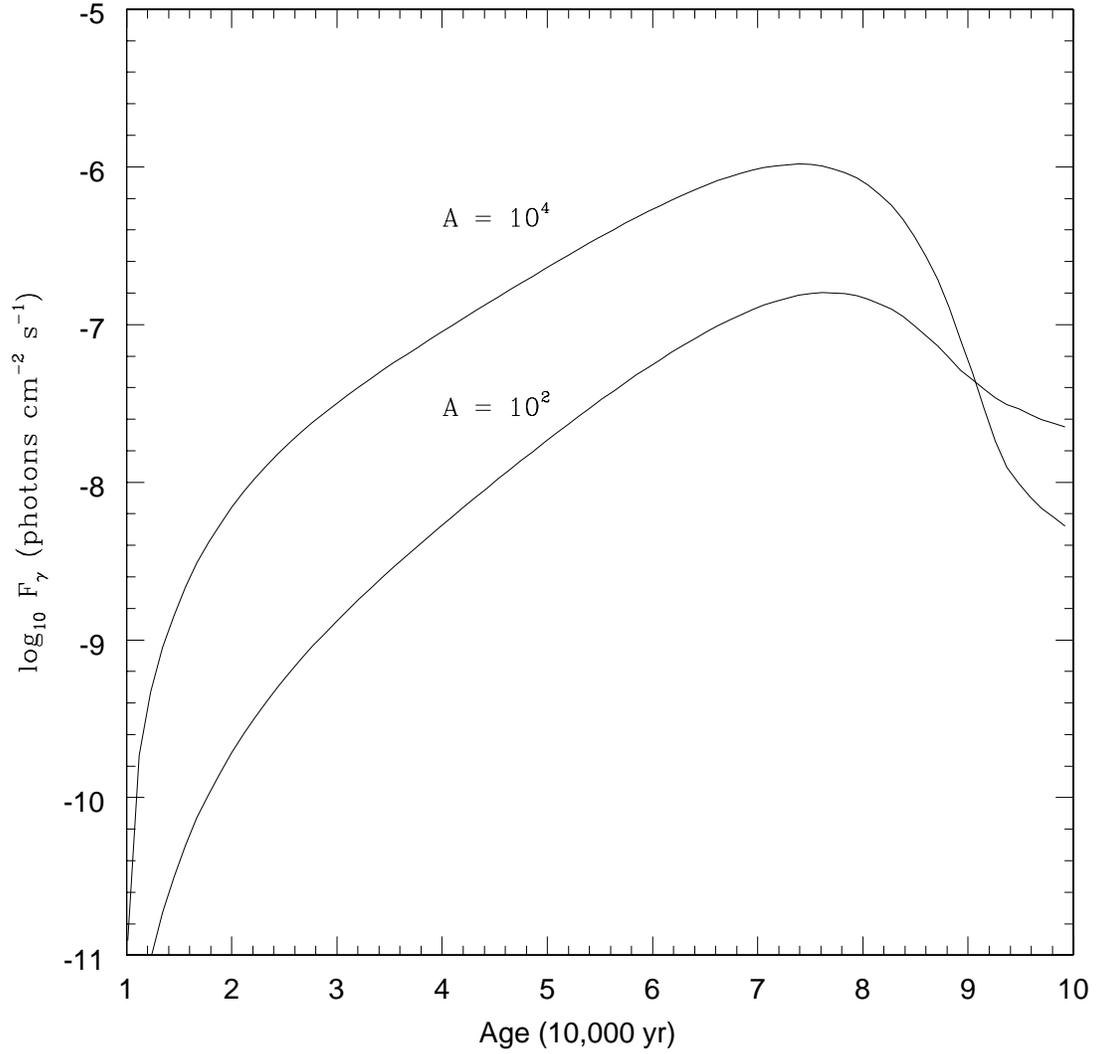}
\end{turn}}
\caption{The flux (as observed at Earth) of the total number of
annihilation photons for both $A=10^2$ and $A=10^4$ as a function
of the age of Sgr A East.  The flux peaks shortly after the
annihilation rate due to charge exchange exceeds the thermalization
rate (see Figure 14).}
\end{figure}

\clearpage
\begin{figure}[thb]\label{fig16.ps}
{\begin{turn}{0}
\epsscale{1.0}
\plotone{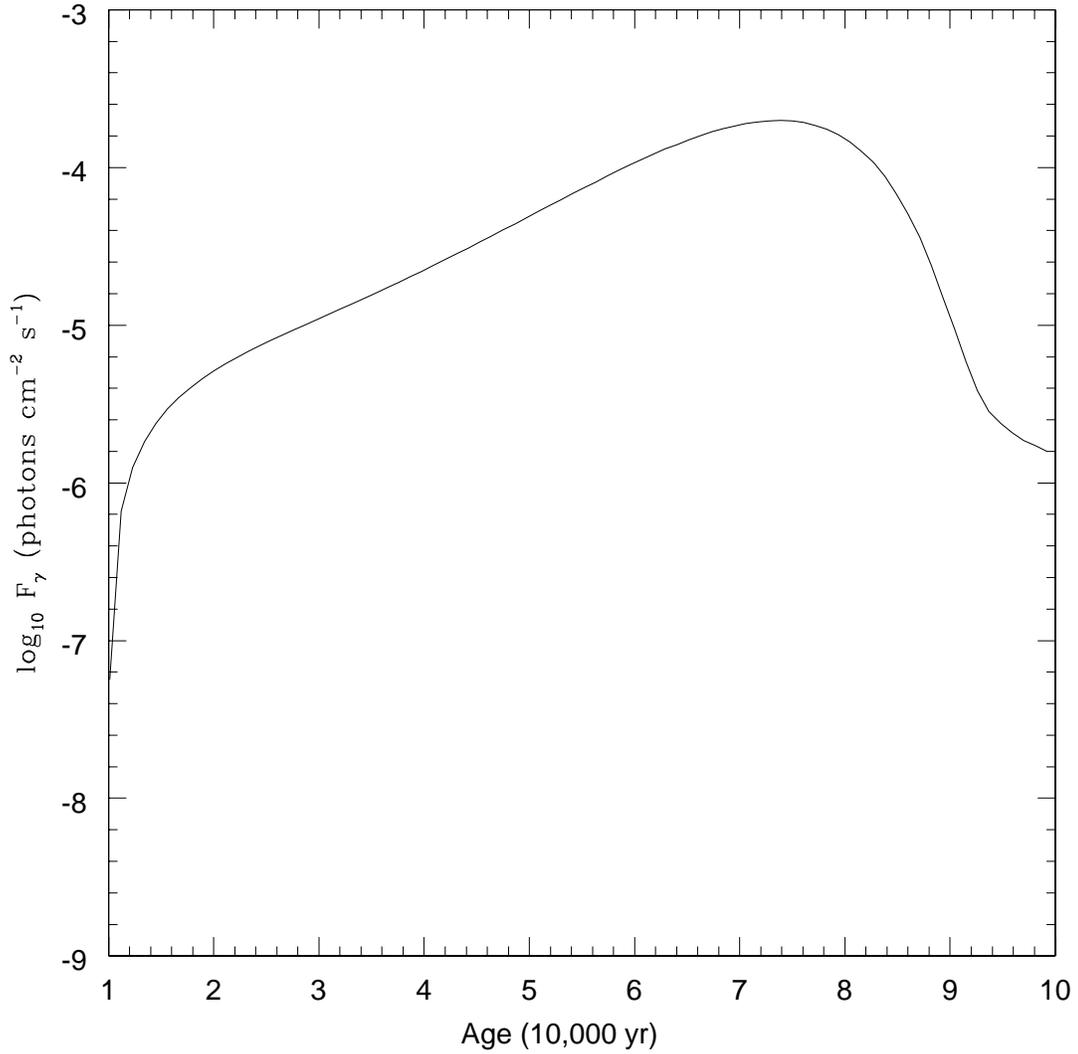}
\end{turn}}
\caption{The flux as observed on Earth of the total number of 
annihilation photons for $A=10^2$ under the {\it ad hoc} assumption
that all injected positrons thermalize on a time scale that is much
shorter than the age of Sgr A East.}
\end{figure}

\end{document}